%% file: MAIN-sigconf.tex
\documentclass[sigconf]{acmart}

\renewcommand\footnotetextcopyrightpermission[1]{} % removes footnote with conference information in first column
\usepackage{booktabs} % For formal tables
\usepackage{soul}
\usepackage{adjustbox}

\usepackage{amsmath}
\usepackage{tabularx}
\newcolumntype{b}{>{\hsize=0.5\hsize}X}

\begin{document}
% \title[Opinion]{Opinion: On the importance of attack detection in ICS as compared to Anomaly detection OR Differentiating attack from anomalies or faults in ICS}

\title[Opinion]{Revisiting Anomaly Detection in ICS: Aimed at Segregation of Attacks and Faults}
%\titlenote{Produces the permission block, and
 % copyright information}
%\subtitle{Cyber Physical System Security: Using the Physics from Physical Part for Attack Detection}
%\subtitlenote{The full version of the author's guide is available as
 % \texttt{acmart.pdf} document}

\author{Chuadhry Mujeeb Ahmed, Jay Prakash and Jianying Zhou \\ (chuadhry,jay\_prakash)@mymail.sutd.edu.sg, jianying\_zhou@sutd.edu.sg \\ Singapore University of Technology and Design}

%\author{Chuadhry Mujeeb Ahmed}
%
% %\authornote{Dr.~Trovato insisted his name be first.}
% %\orcid{1234-5678-9012}
% \affiliation{%
%   \institution{Singapore University of Technology and Design}
%   %\streetaddress{P.O. Box 1212}
%   %\city{Dublin}
%   \state{Singapore}
%   %\postcode{43017-6221}
% }
% \email{chuadhry@mymail.sutd.edu.sg}
%
% \author{Jay Prakash}
%% \authornote{The secretary disavows any knowledge of this author's actions.}
% \affiliation{%
%   \institution{Singapore University of Technology and Design}
%  % \streetaddress{P.O. Box 1212}
%   %\city{Dublin}
%   \state{Singapore}
%   %\postcode{43017-6221}
% }
% \email{jay\_prakash@mymail.sutd.edu.sg}
%
% \author{Jianying Zhou}
% %\authornote{This author is the
%   %one who did all the really hard work.}
% \affiliation{%
%   \institution{Singapore University of Technology and Design}
%   %\streetaddress{1 Th{\o}rv{\"a}ld Circle}
%   %\city{Hekla}
%   \country{Singapore}}
% \email{jianying\_zhou@sutd.edu.sg}

% \renewcommand{\shortauthors}{B. Trovato et al.}

\begin{abstract}
%The domain
In an Industrial Control System (ICS), its complex network of sensors, actuators and controllers have raised security concerns for critical infrastructures and industrial production units. This opinion paper strives to initiate discussion on the design algorithms which can segregate attacks from faults. Most of the proposed anomaly detection mechanisms are not able to differentiate between an attack and an anomaly due to a fault. We argue on the need of solving this important problem form our experiences in CPS security research. First, we motivate using analysis of studies and interviews though economical and psychological aspects. Then main challenges are highlighted. Further, we propose multiple directions of approach with suitable reasoning and examples from ICS systems.
%The Solution

%In this paper, we proposed a technique \emph{Process Skew} that uses the small deviations in the ICS process (herein called as a process fingerprint) for anomaly detection. The process fingerprint is a function of noise in sensor measurements due to the process fluctuations. Such a fingerprint is unique to a process due to the intrinsic operational constraints of the physical process, and is hard to be forged even for a powerful attacker knowing the process operation. We validated the proposed scheme using the data from a real-world water treatment testbed. Our results show that we can effectively identify a process based on its fingerprint, and detect process anomaly with a very low false-positive rate. 

%To tackle a powerful attacker which can learn and imitate the transient behavior, a system model is created to capture the transients of the process. To compromise this an attacker need to know the noise in the transient state of the process which raises the bar for attacker. 
%The Results
\end{abstract}



%, sensor attacks, sensor security}

\maketitle

\input{body.tex}

\bibliographystyle{ACM-Reference-Format}
\bibliography{time_constant}

\end{document}

%% file: body.tex
\section{Introduction}
Industrial Control Systems (ICS) are found in modern critical infrastructure (CI) such as the electric power grid and water treatment plants. The primary role of an ICS is to control the underlying processes in a CI. Such controls are facilitated through the use of computing and communication elements such as Programmable Logic Controllers (PLCs) and  Supervisory Control and Data Acquisition systems (SCADA), and communications networks. The PLCs receive data from sensors, compute control actions, and send these over to the actuators for effecting control over the process. The SCADA workstations are used to exert high level control over the PLCs, and the process, and provide a view into the current process state. Each of these computing elements is vulnerable to cyber and physical attacks as evident from several widely reported successful attempts such as those reported in,\cite{weinbergerStuxnet,ukraineBlackout,germanSteelMill}. Such attacks have demonstrated that while air-gapping a system might be a means to consider securing an ICS, it does not guarantee in keeping attackers from gaining access to the system.

Successful attacks on ICS have led to research to prevent, detect, and react to different forms of cyber attacks. \textit{Anomaly detection} that aims at raising an alert when the controlled process in an ICS moves from its normal to an unexpected, i.e.  {\em anomalous}, state. The challenge with the proposed techniques is that those are not able to distinguish between an anomaly being raised due to an \textit{attack} or a \textit{system fault}.

\section{Problem Identification}
Approaches used in the design of such anomaly detectors fall into two broad categories: \textit{design-centric}\,\cite{adepuMathurTDSC2018} and \textit{data-centric}\,\cite{NoiseMatters_ACSAC2018}. The focus of this position paper is on  the \textit{data-centric} approaches that rely on well-known methods for model creation such as those found in the system identification\,\cite{van1996} and machine learning {literature}\cite{CPS_security_survey2017}. While the use of machine learning to design anomaly detectors becomes attractive with increasing availability of data and advanced computational resources, recent {attempts} \cite{sugumar2019method,CPSweek2016_stealthy_replayATT,adepuMathurTDSC2018,fabioFlorianBulloFrancesco,urbinaGiraldoTipenhauerCardenas} to create anomaly detectors, have concluded that the majority of techniques are not able to distinguish a \textit{fault} from an \textit{attack}. This opinion paper focuses on this challenge with the hope that other researchers will come forward and propose practical solutions to overcome the discussed challenges.
\begin{figure}
	\centering
	\includegraphics[width=0.9\linewidth]{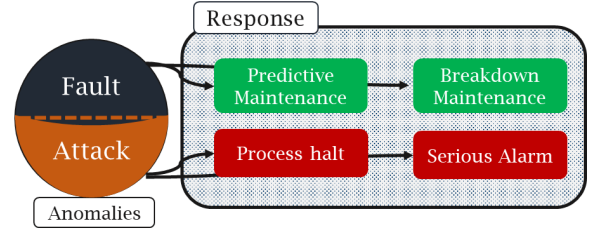}
	\caption{Traditional responses to anomalies: attacks and faults }
	\label{fig:anomlalies}
\end{figure}

$\blacksquare $ \textbf{Relevance of differentiator and correctness:} First, lets investigate why a robust robust differentiator is important. We interviewed plant operators and cyber-security researchers with following questions, as part of a survey, to begin with:
\begin{itemize}
\item Do you have measures to identify between \textit{attack} and \textbf{fault}?

\item If you have measures to identify an attack, what is the typical response to it, i.e., when an alarm is raised for potential attack what is the system and the operator configured to do?
\item Will the response be different if the algorithm told it was a system fault?

\end{itemize}

We interviewed 19 candidates which included researchers at state-of-art test beds like SWaT \cite{swat2016}, EPIC \cite{siddiqi2018practical} and WADI \cite{wadi2017}, experts in ICS security and engineers at industrial production plants for steel and water.
18 participants said that they have not adopted any measure so far or have not come across even a near-robust algorithm to differentiate between the two versions of anomalies.

Based on responses to aforementioned questions, we could confirm that both the system's and the people's behavior vary a lot in the contexts of attacks and fault alarms. We sketch the findings in Fig. \ref{fig:anomlalies}. A CI plant may go in shutdown for hours under identification of an attack. A correct classification, an attack identified as an attack and a fault as a fault would save loss of lives and revenues. But incorrect and cross classifications would be detrimental for the system. A fault raised as attack has high psychological impacts, invites panic and responses of higher magnitudes. Hence, if anomaly detectors are deployed, faults in sensors or actuators might lead to shutdowns of whole units and incur huge economic losses. Similarly, an attack identified as a fault would undermine the losses and a repetitions would discourage adoptions of such detectors. We note that psychologically an attack is dealt with much higher alertness and hardness in response strategy. 

Our participants did acknowledge the relevance which can be inferred from the responses like:
\begin{itemize}
    \item \textit{"Yes, no need to shut the whole system} under assumption of attack when a fault might have occurred".
    \item \textit{"Yes, if an algorithm can identify it to be a system fault immediately, it will reduce the amount of time required during the identification process. True attacks can be immediately treated as attacks and response time will be much shorter."}
    \item \textit{"Yes, the incidents should be handled differently."
}
\item \textit{"... But the incident response may have different outcome if it isn't an attack. Such as forensics and developing control to handle such attacks may not be performed."}
\end{itemize}

Five of the participants reported to be working on response and mitigation strategy design and did suggest that differentiating strategy would save time and facilitate narrowing down focus and search spaces. Participants working in production industries reported of following traditional methods of \textit{predictive maintenance} in case of anomalies i.e., deviations of variables from history and process physics Fig. \ref{fig:anomlalies}. The process still keeps running in this case. \textit{ Breakdown maintenance} is conducted in case there are serious faults. In this case either spare parts are used or redundant process chains are activated. It is to be noted that there are dedicated employees under \textit{ central maintenance team} for handling such faults which have been maturing over the decades but intelligent cyber attackers might hide their attacks inside such checks. Also the frequency of faults and predictive maintenance are fairly high.  On the other side, a cyber attack leads to process halt and shutdowns of plants like in recent case with Renault, \cite{renault}. And to make matter worse, cyber attacks against industrial targets have been growing rapidly \cite{rise} as well. As noted before, any cross-error would invite economic and opportunity losses. Hence a segregation is necessary and very timely.

%\hl{Highlight how big the trouble is,economic physical losses}

%\hl{Psychological differences}

%\hl{ A set diagram}

\section{Why is the segregation difficult?}

\begin{figure}
    \centering[h]
    \includegraphics[scale=0.11]{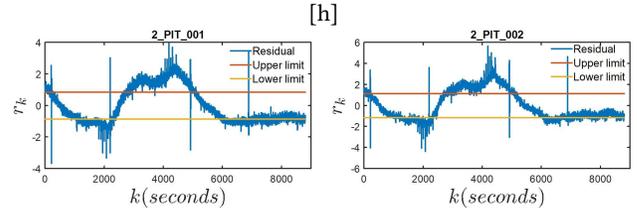}
    \caption{It is hard to distinguish between an attack (on the left) and a fault (on the right) for two different sensors in a water distribution testbed.}
    \label{fig:leak_exp_wadi}
\end{figure}

We and past research works like \cite{sugumar2019method,CPSweek2016_stealthy_replayATT,adepuMathurTDSC2018,Asiaccs2018_mujeeb_noiseprint,fabioFlorianBulloFrancesco,urbinaGiraldoTipenhauerCardenas,John_ACSAC2018} note that it is fairly difficult to differentiate between these two vectors of anomalies. As an example consider Fig.~\ref{fig:leak_exp_wadi} that shows data from two different pressure sensors from the water distribution testbed. On the left hand side, a water leakage attack was executed while on the right hand side a fault similar to leakage happened but for a statistical detector both appear to be the same. Few of the challenges are summarized herein:
%\hl{Why other people/research cannot differentiate}
\\

% \textcolor{red}{May be highlight some of our experiences...like invariants....noisense......model based....}
$\blacksquare$ \textbf{Challenge 1. Not modeling properties of faults and attacks:} Most of the related works would look at the consequences of an attack or fault rather than looking through the properties of attack or fault themselves. For example, a fault is usually random and results in abrupt change for a short period of time. On the other hand an attack is properly planed and is executed for a longer time to do substantial damages. Missing on opportunities of differential properties adds to hardness of the problem. \\ 

$\blacksquare$ \textbf{Challenge 2. Unknown attacks and faults:} The base assumption for an anomaly detection method is that there would always be unknown attacks, therefore it is not possible to use blacklisting or white-listing as an effective method~\cite{sommer2010paxson_anomaly_detection_ML}. Therefore, we rely on behavioral methods which simply look at the end result of anomalous behavior and raise an alarm. The challenge is that it is not possible to create a model for each possible fault that is out there in the wild. 
\newline

$\blacksquare$ \textbf{Challenge 3. Lack of hybrid models:}
There are no precise hybrid models for cyber and physical domains. It is not clear how to come up and use the information from independent and orthogonal spaces through an analytical lense. Most of the anomaly detection methods are implemented either in cyber layer or in the physical layer. There is no one size fit all techniques due to a wide variety of ICS.  

\section{Suggested Directions of Approach}

There are a lot of efforts for detecting sensor attacks but most can not distinguish between an attack and a faulty sensor measurement. An attempt was made recently~\cite{park2015sensor_transientFaults} to model and detect transient faults. 
 It models a transient fault for each sensor and an algorithm is designed to detect and identify attacks in the presence of transient faults. This approach can detect transient faults (e.g., a GPS reporting faulty readings inside a tunnel) but not the permanent faults/attacks, e.g.,  DDoS attack or cutting wire of a sensor. Also, this work does not consider a stealthy attacker trying to imitate a  transient fault. They have also assumed multiple sensors for the same physical state variable. They also assume an abstract sensor model where the sensor reports an interval of readings e.g., a set rather than one value. Therefore, it is highly desirable to come up with novel methods which could differentiate between a fault and an attack by considering a realistic threat model. In the following, we present a few proposals that also highlight open problems.

% \hl{should start with the most convincing one first?}

\subsection{Combining Process and Network Layer Detectors}
The studies involving both the network and the process layer are    rare~\cite{CPS_security_survey2017}. We propose to use data from both the network layer and the process layer. For example, if a sensor data is compromised as a man in the middle attack (MiTM)~\cite{NoiseMatters_ACSAC2018} then the resultant process state might have equally resulted from a fault but by looking at the MiTM traffic, it would be certainly possible to point out an attack.

\subsection{Relation Between Size, Detection Time and Time to Damage of faults and attacks}
Fault/Attack size means how much is the sudden change in the physical state variables. For example, consider an initial state ($S_i$) before attack and state transition to $S_a$ after an attack, where $|S_a| >>> |S_i|$, such an instantaneous change would be considered of a large size. Detection time is a measure of  how fast one can detect the attack whereas time to damage depends on the process itself, e.g., for a fluid storage tank depending on the capacity it could take long time to overflow or underflow attack. While in electric grid a sudden surge can instantly damage the physical system. We can study relationship between variables to segregate between two.

%Methods which are, for example, designed for the detection of abrupt faults may be not suitable for slowly developing faults, and vice versa. Therefore several methods may be needed in parallel, also to build up redundant fault detection systems. 

\subsection{Virtual Sensors/Digital Twin}
We can exploit the idea of a digital twin or model virtual sensors. Since the digital system does not have any real sensors, it could not get faulty and any manipulation must be a result of some attack. The key assumption is that an attacker must attack that virtual/digital twin.

\subsection{Signature based Attack Detection}
Attack signatures can be collected by designing an attacker's intentions and strategy. However, a limitation of this method is that it would not be able to detect unseen attack patterns.

\subsection{Mode Shift Across Attack and Fault}
The idea is to figure out how the device's mode transition occurs during a fault. Faults would be random, for example, failing of a sensor or actuator. Whereas, attacks would exhibit a mode shift for several devices at the same time. For example, to attack a process plant, an attacker would compromise the sensor as well as actuator's mode at the same time to hide itself.

\subsection{Simulate Failures and Faults for Known Device}
Attacks are unknown before an attack has taken place whereas fault data could be generated. The idea is to collect data for the (known) faulty behavior to create fault models. If we have profiles for normal data as well as faulty data, then it should be possible to distinguish attacks from faults.

\subsection{ Exploiting Asymmetry between Correlation and Causation}
We argue that correlated failures should be taken with a pinch of salt. If a sensor fails or an actuator fails it would have an impact on another associated device. For example, if a motorized valve at the inlet of a tank fails then the flow sensor is also affected but if the flow sensor fails then the actuator (motorized valve) is not affected. Therefore, to attack the flow sensor an attacker has to attack both the flow meter as well as the actuator but there is no such requirement for a fault.  % however if level sensor fails actuator might be affected. 

% \subsection{How is change in states differ during an attack and fault or even in normal operation?}
% \textcolor{red}{I think this is similar to point 1 and comment it if need some text can bring to point 1...what do you think???}
% This is to state that the change in the physical process state variable would be abrupt in case of fault as compared to an attack. But the challenge is that some of the attacks could be fault like which just want to compromise the safety of the physical system without paying much attention to stay hidden. 

\subsection{Detection Latency based Method}
An interesting observation from previous research~\cite{Ahmed_AsiaCCS2017_stealthyAtt,park2015sensor_transientFaults} shows that sophisticated attacks (takes more time) and fault like bias injection attacks (detected instantaneously) takes different amount of time. The hypothesis is that the fault would be a sudden change. However, a persistent attacker would stay for a longer period of time to do substantial damage. 

%\textcolor{red}{Above was my idea but similar Copied from the PAjic PAPER: "Due to their short duration, however, transient faults should not be considered as a security threat to the system. In contrast, permanent faults are sensor defects that persist for a longer period of time and may seriously affect the system’s operation. For instance, a sensor may suffer physical damage that introduces a permanent bias in its measurements. In such a scenario, unless the fault can be corrected for in the software, the system would benefit from discardingthis sensor altogether."}

% \textcolor{red}{Following 2 does not amke much sense to me}
% \subsection{On active detection e.g. watermarking}

% active detection methods like challenge response should be able to figure out between an attack and fault...right?

% \subsection{Accurate fault models and active defense}
% Provided accurate fault models aided by watermarking approaches can be a promising solution. 

\subsection{Fault Time Constant}
Fault's time constant would be different than that of attack. Inspired from the research in fault detection in ICS~\cite{abrupt_fault_time_constant_samara2008statistical}, it is hypothesized that the faults are more abrupt and random and hence their time constant is much smaller than the normal process profile. However, we assume that an attacker would try to imitate as close to the process as possible, therefore if modeled properly faults could be distinguished.

\subsection{Redundancy} Using the redundant sensors for the same physical state variable can help in fault and attack isolation provided that not all are attacked at the same time~\cite{park2015sensor_transientFaults}. 

% \section{Critics/Related Work}

% Table 

\section{Conclusions}
Through this opinion paper we aim to kindle interest of ICS community towards deeper nuances of anomaly detection and analysis. We assert the relevance of differentiating between faults and attacks in cyber-physical systems and try to motivate from both economical and psychological lenses which are further corroborated by interviews with researchers and industry managers. We build up on top of the core challenges in segregating the two forms of anomalies and propose multiple directions of approach. This work shall be followed by rigorous research in most of the outlined directions in collaborations with research institutes and universities hosting state-of-art CPS test beds.